\documentclass[aps,pra,showpacs,twocolumn,groupedaddress]{revtex4}
\usepackage{amsmath}
\usepackage{amssymb}
\usepackage{graphics}
\usepackage[dvips]{graphicx}
\usepackage{graphicx}
\usepackage{color}

\newcommand \beq{\begin{eqnarray}}
\newcommand \eeq{\end{eqnarray}}
\def\simge{\mathrel{%
       \rlap{\raise 0.511ex \hbox{$>$}}{\lower 0.511ex \hbox{$\sim$}}}}
\def\simle{\mathrel{
       \rlap{\raise 0.511ex \hbox{$<$}}{\lower 0.511ex \hbox{$\sim$}}}}

\newcommand{\rr}{\mathbf{r}}
\newcommand{\kk}{\mathbf{k}}

\begin{document}
\title{Short-range correlations and entropy in ultracold atomic Fermi gases}
\author{Zhenhua Yu,$^a$ Georg M. Bruun,$^a$ and Gordon Baym$^{a,b}$}
\affiliation{ $^a$The Niels Bohr International Academy, The Niels
Bohr Institute, Blegdamsvej 17, DK-2100 Copenhagen \O,
 Denmark\\
$^b$Department of Physics, University of Illinois, 1110
  W. Green Street, Urbana, IL 61801
}

\date{\today}

\begin{abstract}
We relate short-range correlations in ultracold atomic Fermi gases
to the entropy of the system over the entire temperature, $T$, vs.
coupling strength, $-1/k_Fa$, plane.  In the low temperature limit
the entropy is dominated by phonon excitations and the correlations
increase as $T^4$.  In the BEC limit, we calculate a boson model
within the Bogoliubov approximation to show explicitly how phonons
enhance the fermion correlations.  In the high temperature limit, we
show from the virial expansion that the correlations decrease as
$1/T$. The correlations therefore reach a maximum at a finite
temperature. We infer the general structure of the isentropes of the
Fermi gas in the $T,-1/k_Fa$ plane, and the temperature dependence
of the correlations in the unitary, BEC, and BCS limits.  Our
results compare well with measurements of the correlations via
photoassociation experiments at higher temperatures.
\end{abstract}

\maketitle

\section{Introduction}

The thermodynamics of an ultracold atomic gas interacting via short
range interactions is encoded in the short range two-body
correlations of the particles.  Such connections between short range
correlations and physical properties of systems with short range
interactions have been long recognized, e.g., Ref.~\cite{mfisher},
and have been employed recently to study the rf spectra of paired
Fermi gases \cite{bpyz,zwerger,zhang}. Tan has derived a number of
remarkable relations between the properties of ultracold atomic
gases, e.g., the ground state energy and their short range
correlations \cite{shina_ann1} (see also \cite{Braaten}). These
links open new perspectives to examine the many particle physics of
systems with short range interactions.

By using Feshbach resonances to tune the scattering length $a$, one
can study atomic Fermi gases through the crossover from a weakly
coupled BCS superfluid to a weakly coupled Bose-Einstein condensate
(BEC) of molecules.  In the strongly correlated unitarity regime,
$|a|\to\infty$, the many-body correlations are highly non-trivial
and one has to resort to Monte-Carlo calculations to obtain
controlled results~\cite{giorgini,giorgini_corr}. At non-zero
temperatures $T$, information on the correlations is even more
limited.  We study here how knowledge of the free energy and entropy
can illuminate the nature of the finite temperature short range two
particle correlations. The two-body correlation function in a two
component Fermi gas has the general structure \beq
\langle\psi^\dagger_1(\rr')\psi^\dagger_2(0)\psi_2(0)\psi_1(\rr)\rangle
= \sum_i \gamma_i \phi_i(\rr) \phi_i^*(\rr'), \label{gamma} \eeq as
one sees by regarding the correlation function as a Hermitean
operator in $\rr,\rr'$. For $r\ll d$, the interparticle spacing, the
functions $\phi_i$, essentially s-wave Jastrow factors in the
many-body wavefunction at short interparticle distance, are
determined by two-body physics, and outside the range of the
potential, $r_0$, at temperatures of interest, are $\sim \sin(kr
+\delta)/r = \sin\delta \,\,\chi(r)/r$, where $\chi(r) = 1-r/a$,
$\delta$ is the s-wave phase shift, and $\cot\delta = -1/ka$, where
$a$ is the s-wave scattering length.  Thus
\begin{align}
\langle\psi^\dagger_1(\rr)\psi^\dagger_2(0)\psi_2(0)\psi_1(\rr)\rangle
=C\left(\frac{\chi(r)}{r}\right)^2 = C\left(\frac1r-
\frac1a\right)^2. \label{C1}
\end{align}
The correlation (or contact) strength $C$, which is determined by many-body physics, can be
 measured directly in photoassociation experiments \cite{randy,castin}.

Here we examine the temperature dependence of the correlations by
calculating  $C(T)$ from the finite temperature thermodynamics.
Quite generally, the short range correlations are related to the
free energy density $f={\rm Tr}e^{-H/T}/\Omega$ (where $\Omega$ is
the volume of the system) by \cite{zhang}:
\begin{align}
C=-\frac{m}{4\pi}\frac{\partial f}{\partial a^{-1}}.
\label{e:c1}
\end{align}
Differentiating with respect to the temperature, $T$, we find
\begin{align}
\frac{\partial C}{\partial T}= -\frac{m}{4\pi}\frac{\partial}{\partial T}\frac{\partial f}{\partial a^{-1}} =\frac{m}{4\pi}\frac{\partial
s}{\partial a^{-1}},
\label{e:c2}
\end{align}
thus linking the variation of $C$ with temperature to the coupling
constant dependence of the entropy density $s$. The connection is
illustrated below in Fig. \ref{Cplot} which shows the temperature
dependence of $C$, and the related Fig. \ref{isentrope_plot}, which
shows the isentropes in the $T,-1/k_Fa$ plane.

We focus on the unitarity, BEC, and BCS limits and provide
controlled results for $C$ for both high and low $T$.  Using the
virial expansion, we show that the correlations decrease with
$C\propto n^2/T$ in the high temperature limit, where $n$ is the
density.  From the relation (\ref{e:c2}) between the $T$ dependence
of $C$ and the entropy density, we show that $C(T)$ \emph{increases}
with temperature as $T^4$ for low $T$, owing to the phonon
contribution to the entropy.   In the BEC limit, we assume a bosonic
model to relate the bosonic and fermionic correlations; a
calculation within the Bogoliubov approximation illustrates how
phonons enhance the fermion correlations at low temperature.
Combining the behavior of $C$ in the high and low $T$ limits, we
infer that $C$ reaches a maximum value at a temperature  $T_{\rm
max}>0$.  This intriguing behavior of $C(T)$ is detectable in
photoassociation experiments near unitarity.

\section{Correlation function and free energy}
We consider a homogenous two-component Fermi gas described by the Hamiltonian ($\hbar=1$ throughout)
\begin{align}
H=& \int d^3r \sum_{i=1,2}\frac{1}{2m} \nabla
 \psi_i^\dagger(\rr)\cdot\nabla\psi_i(\rr)
 \nonumber\\
 &+\int d^3r \int d^3 r' V(|\rr-\rr'|)
 \psi_1^\dagger(\rr)\psi_2^\dagger(\rr') \psi_2(\rr') \psi_1(\rr),
 \label{e:ham}
\end{align}
where the $\psi_i$ are the field operators for fermions in internal state $|i\rangle$. We take the range $r_0$ of the interaction $V(r)$ to be
 short compared to the interparticle spacing $d$.

We first present an elementary derivation of the relation (\ref{e:c1}) between the two-body correlation function and the free energy density, using the device of scaling the potential $V(r)$ by a factor $\lambda$ in  (\ref{e:ham}); $V(r)\to \lambda V(r)$, letting $\lambda \to 1$ in the end.  The correlation function for given $\lambda$ has the structure of  (\ref{C1}) with $\chi$, $C$ and $a$ functions of $\lambda$,
For $r \ll d$,  the function $\chi_\lambda$ satisfies the low energy limit of the two-body Schr\"odinger equation \cite{leggettbook,zhang}
\begin{align}
\left(-\frac1m\frac{d^2}{dr^2}+\lambda V(r)\right)\chi_\lambda(r)=0.
\label{lambda}
\end{align}
At short distances, with $r>r_0$,  $\chi_\lambda(r)=1-r/a_\lambda$; for convenience, we choose the normalization constant to be unity. Differentiating  (\ref{lambda}) with respect to $\lambda$, multiplying by $\chi_\lambda$, integrating with respect to $r$ from 0 to $r_c$, and eliminating the term $\int dr \chi_\lambda V \partial \chi_\lambda/\partial \lambda$ using  (\ref{lambda}), we find
\begin{align}
\int_0^\infty dr\; V(r)\chi^2_\lambda(r)\frac{\partial\lambda}{\partial a_{\lambda}^{-1}}=-\frac1m\label{dlda},
\end{align}
where we use vanishing of the s-wave function $\chi_\lambda$ at the origin, and extend the upper bound of the integral from $r_c$ to infinity since $V(r)$ is nonzero only for $r<r_0$.
On the other hand, from the Feynman-Hellmann theorem, the variation of the free energy density with coupling strength is
 \begin{align}
\frac{\partial f}{\partial a_{\lambda}^{-1}}=&\frac{\partial f}{\partial \lambda}
\frac{\partial\lambda}{\partial a_{\lambda}^{-1}}=\frac1\Omega\left\langle\frac{\partial H}{\partial \lambda}\right\rangle\frac{\partial\lambda}{\partial a_{\lambda}^{-1}}\nonumber\\
=&\int d^3 r\; V(r)\langle\psi^\dagger_1(\rr)\psi^\dagger_2(0)\psi_2(0)\psi_1(\rr)\rangle_\lambda\frac{\partial\lambda}{\partial a_{\lambda}^{-1}}\nonumber\\
=&4\pi C_\lambda \int_0^\infty dr\; V(r)\chi^2_\lambda(r)\frac{\partial\lambda}{\partial a_{\lambda}^{-1}}.\label{dfda}
\end{align}
Combining  (\ref{dlda}) and (\ref{dfda}),  and setting $\lambda=1$,
we arrive at the desired relation (\ref{e:c1}).

We review the application of (\ref{e:c1}) to a homogeneous equally
populated two-component Fermi gas of volume $\Omega$, at $T=0$
\cite{castin}.  The ground state energy per particle $E/2N$ can be
written
 as $(3/5)(1+\beta)E_F$, where $\beta$ is a
function of the dimensionless quantity $\xi=-1/k_F a$. The Fermi
momentum $k_F$ is defined in terms of the single component density
$n=N/\Omega$ by $n=k_F^3/6\pi^2$, and the Fermi energy by
$E_F=k_F^2/2m$. Equation (\ref{e:c1}) implies that $
C(T=0)=(k_F^4/40\pi^3)\partial\beta/\partial\xi$. The function
$\beta$ has been calculated by quantum Monte Carlo simulations near
unitarity ($1/a\to 0$); at unitarity $\partial\beta/\partial
\xi\approx 0.9$~\cite{giorgini}, a result which agrees well with the
value $C=2.7 (k_F^4/36\pi^4)$ extracted directly from the
correlation function calculated by quantum Monte Carlo methods
\cite{giorgini_corr}. In the BCS limit, $-1/k_F a\gg1$, a
perturbative expansion in $k_F a$ gives \cite{huang,lee,galitskii}
\begin{align}
 \frac{E}{2N}=\frac35 E_F\left[1+\frac{10}{9\pi}k_Fa+\dots\right].
 \label{betabcs}
\end{align}
Corrections to $E$ due to pairing are exponentially small in this
limit. Keeping the leading term yields $C=a^2n^2$. In the BEC limit,
$1/k_F a\gg1$, fermions form molecules and~\cite{huang,lee}
\begin{gather}
\frac{E}{2N}=\frac{E_b}{2}+ \frac{E_F k_F
a_m}{6\pi}\left(1+\frac{128}{15\sqrt{6\pi^3}}(k_Fa_m)^{3/2}+\dots\right),\label{betabec}
\end{gather}
with $E_b=-1/ma^2$ the molecular binding energy, and $a_m$ the
scattering length between molecules. The second term is the
Hartree-Fock mean field energy and the third the
nonperturbative Lee-Yang correction \cite{lee}. Few body
calculations give $a_m=0.6a$ \cite{petrov}.  In the BEC limit, $C=
n/2\pi a$ to leading order. The divergent behavior, as $a\to0$,
arises from the normalization of the molecular wave function,
$\chi/r\sim e^{-r/a}/ra^{1/2}$.

\section{Low temperature}

We now consider the temperature dependence of $C$ for $T$ well below the superfluid transition temperature $T_c$,
in which regime phonon excitations dominate.
The contribution of the phonons to the entropy density is
\begin{align}
s_{\rm phonon}=\frac{2\pi^2}{45} \left(\frac T{c_s}\right)^3,
\label{sph}
\end{align}
where the zero temperature sound velocity $c_s$ is given by
$mc_s^2=n\partial \mu/\partial n$, with $\mu$ is the chemical
potential. The change of $C$ to leading order for $T\ll T_F=E_F$
($k_B=1$) is thus
\begin{align}
\delta C=C(T)-C(0)=-\frac{\pi m}{120}\left(\frac T{c_s}\right)^4\frac{\partial c_s}{\partial a^{-1}}.\label{clow}
\end{align}
The zero temperature sound velocity
appears to increase monotonically from the BEC side to the BCS
side~\cite{giorgini,sound,Combescot}, so that the short range pair
correlations as parametrized via $C$  \emph{increase} from zero
temperature as $T^4$.  At $T=0$,
$\mu=E_F\left(1+\beta-\beta'\xi/5\right)$, so that
\begin{align}
\left(\frac{c_s}{v_F/\sqrt3}\right)^2=1+\beta-\frac35\beta'\xi+\frac1{10}\beta''\xi^2,
\label{vs}
\end{align}
with $\beta'=\partial\beta/\partial\xi$, etc. Recent measurements of $c_s$ are in good agreement with (\ref{vs}) combined with Monte Carlo
results for $\beta$ around unitarity~\cite{thomas}.

Equation (\ref{betabcs}) implies that in the BCS limit,
$c_s^2=(v_F^2/3)(1+2k_Fa/\pi)$, and thus at low $T$,
\begin{align}
C(T)=n^2a^2\left[1+\frac{9\sqrt{3}\pi^4}{160}\left(\frac{T}{T_F}\right)^4\right].
\label{BCSlimit}
\end{align}
For $T_c<T\ll T_F$, the weakly attractive Fermi gas
can be described by Landau Fermi liquid theory. The entropy density is given by \cite{gordon_chris}
$s_f=\frac13 m^*k_FT$ with $m^*=m[1+8(7\log2-1)(k_F a)^2/15\pi^2]$  the effective mass of the quasiparticles \cite{lifshitz}.
From  (\ref{e:c2}), we then have the temperature dependence
\begin{align}
\delta C(T)/n^2a^2=-\frac\pi 5(7\log2-1)k_F a\left(\frac T{T_F}\right)^2>0.\label{cfl}
\end{align}

In the BEC limit, the Bogoliubov sound speed is $c_s=\sqrt{ng_m/M}$ with $g_m=4\pi a_m/M$ and $M=2m$~\cite{PethickSmith}. Thus
\begin{align}
C(T)=& \frac{n}{2\pi a}\Big[1 +\left(\frac{\partial a_m}{\partial a}\right)\frac{(k_Fa)^3}{12\pi}\Big(1+\frac{64}{3\sqrt{6\pi^3}}(k_F a_m)^{3/2}\nonumber\\
&+\frac{ (486\pi^{13})^{1/2}}{40 (k_Fa)^{5/2}}
\left(\frac{T}{T_F}\right)^4\Big)
  \Big].
\label{BEClimit}
\end{align}
At unitarity,
\begin{equation}
C(T)=\frac{k_F^4\beta'}{40\pi^3}\left[1+\frac{\sqrt{3}\pi^4}{80(1+\beta)^{5/2}}\left(\frac{T}{T_F}\right)^4\right].
\label{Unitaritylimit}
\end{equation}

\section{correlations in the BEC limit}
The universal behavior $\delta C\sim T^4$ as $T\to0$ shows that thermally excited phonons enhance the
fermion pair correlations. In the BEC limit, the enhancement of the correlations can be understood directly in terms of a gas of molecules interacting
with a short range potential with a  scattering length $a_m>0$.  The boson correlation function is similar in structure to that for fermions:
\begin{align}
&\lim_{r\to0}\langle\phi^\dagger(\rr)\phi^\dagger(0)\phi(0)\phi(\rr)\rangle=C_m\left(\frac1 r-\frac 1{a_m}\right)^2,
\label{mc}
\end{align}
where the $\phi(\rr)$ are the bosonic field operators.  As for fermions,
\begin{align}
 C_m=-\frac{M}{2\pi}\frac{\partial f_m}{\partial a_m^{-1}},
 \label{cm}
\end{align}
with $f_m$ the free energy density for the bosonic molecules.  The factor of 2 difference from (\ref{e:c1}) is due to the factor 1/2 in the interaction energy for identical bosons written in terms of the $\phi$.

Since the low temperature physics described in terms of the fundamental fermions or bosonic molecules must be the same, the leading temperature variation $\delta f=f(T)-f(0)$ is equal to $\delta f_m=f_m(T)-f_m(0)$. Therefore
\begin{align}
\delta C=-\frac m{4\pi} \frac{\partial (\delta f)}{\partial a^{-1}}
=&\, -\frac{m}{4\pi}\frac{\partial (\delta f_m)}{\partial a_m^{-1}}\frac{\partial a_m^{-1}}{\partial a^{-1}}\nonumber\\
=&\,\frac m{2M}\frac{\partial a_m^{-1}}{\partial a^{-1}}\delta C_m,\label{ccm}
\end{align}
with $\delta C_m=C_m(T)-C_m(0)$.  We calculate the boson correlation
function within the Bogoliubov approximation using a pseudopotential
$V_m(r)=g_m\delta(\rr)$.   We divide the field operator
$\phi(\rr)=\phi_0+\delta\phi(\rr)$ into a condensate part $\phi_0$
and a fluctuation $\delta\phi$, assumed to be small.  To
second order in the fluctuations, at small $r$,
\begin{align}
&\langle\phi^\dagger(\rr)\phi^\dagger(0)\phi(0)\phi(\rr)\rangle\nonumber\\
&=n_0^2+2n_0[2\langle\delta\phi^\dagger(0)\delta\phi(0)\rangle
+\langle\delta\phi^\dagger(\rr)\delta\phi^\dagger(0)\rangle],
\label{phi4}
\end{align}
where
\begin{align}
&\langle\delta\phi(0)^\dagger\delta\phi(0)\rangle=\int\frac{d^3k}{(2\pi)^3} [v_k^2+(u_k^2+v_k^2)\langle\alpha_k^\dagger\alpha_k\rangle],\\
&\lim_{r\to0}\langle\delta\phi^\dagger(\rr)\delta\phi^\dagger(0)\rangle=-\int\frac{d^3k}{(2\pi)^3} u_kv_k[e^{i\kk\cdot\rr}+2\langle\alpha^\dagger_k\alpha_k\rangle]\nonumber\\
=&-\frac{n_0 a_m}r-\int\frac{d^3k}{(2\pi)^3}\left(u_kv_k-\frac{Mg_m n_0}{k^2}
  +2u_kv_k\langle\alpha^\dagger_k\alpha_k\rangle\right);
\end{align}
here $v_k^2=(\xi_k/E_k-1)/2$, $u_k^2=(\xi_k/E_k+1)/2$, $\xi_k=k^2/2M+g_m n_0$, and the $\alpha_k$ annihilate phonons with energy dispersion $E_k=\sqrt{\xi_k^2-(g_m n_0)^2}$.
The $1/r$ divergence in $\langle\delta\phi^\dagger(\rr)\delta\phi^\dagger(0)\rangle$ corresponds to the cross term in  (\ref{mc}).
The $\sim 1/r^2$ in  (\ref{mc}) appears from the higher order term $\lim_{r\to0}\langle\delta\phi^\dagger(\rr)\delta\phi^\dagger(0)\rangle \langle\delta\phi(\rr)\delta\phi(0)\rangle =
\lim_{r\to0}|\langle\delta\phi(\rr)\delta\phi(0)\rangle|^2+\ldots$, not included in (\ref{phi4}).

The lowest order Bogoliubov approximation produces the correlation
function (\ref{mc}) to leading order in $(na_m^3)^{1/2}$.  In
(\ref{phi4}), the $r$ independent term is reliable to order
$(na_m^3)^{1/2}$, the term proportional to $1/n^{1/3}r$ to order
$(na_m^3)^{1/3}$, and the term proportional to $1/(n^{1/3}r)^2$ to
order $(na_m^3)^{2/3}$.   To derive the full $C_m(1-a_m/r)^2$
structure to higher order in $(na_m^3)^{1/2}$ requires calculating
terms beyond the simple ones in  (\ref{phi4}), a task we defer.
To extract $C_m$ from the $r$ independent term in  (\ref{phi4})
we write $n_0=n-\langle\delta\phi(0)^\dagger\delta\phi(0)\rangle$.
At $T=0$, $C_m/(na_m)^2=1+(64/3)\left(na_m^3/\pi\right)^{1/2}$,
which agrees with the calculation of $C_m$ from  (\ref{cm}) using
the ground state energy of weakly interacting bosons
[cf. (\ref{betabec})]. The thermally induced change $\delta C_m$
is
\begin{align}
\frac{\delta C_m}{2na_m^2}=&\int\frac{d^3k}{(2\pi)^3}(u_k^2+v_k^2-2u_kv_k)\langle\alpha_k^\dagger\alpha_k\rangle\nonumber\\
=&\int\frac{d^3k}{(2\pi)^3} \frac{k^2}{2E_k}\frac1{e^{E_k/T}-1}=\frac{\pi^2}{60}\frac{T^4}{c_s^{5}},\label{dcm}
\end{align}
where we consider only the phonon contribution ($E_k=c_sk$).
Substituting  (\ref{dcm}) into (\ref{ccm}), we obtain the same result as in (\ref{BEClimit}).

The $T^4$ dependence can be understood in terms of the effect of
thermal sound waves on the eigenstates, $\phi_i(r)$, of the two
particle density matrix (\ref{gamma}). At distances beyond the
interparticle spacing,  Eq.~(\ref{lambda}) for the $\phi$
(with $\lambda=1$) contains potential terms from
the mean field $g_mn$ and its fluctuations.  At fixed normalization,
the latter change the magnitude of $\phi_i(r)$ at short
distances by terms $\sim \langle (\delta n)^2  \rangle \sim T^4$,
which translates into the $T^4$ dependence of $C$ at low
temperature.

\section{High temperature}
In the high temperature limit, $T\gg T_F$, where the fugacity
$z=e^{\mu/T}$ is small, the free energy can be calculated via a
virial expansion.  To second order in $z$, the partition function
for a two-component Fermi gas is given by \cite{landau,jason}
\begin{align}
\log (Z/Z_0)=\frac{2^{3/2}\Omega}{\lambda^3}z^2b_2, \label{virial}
\end{align}
where
\begin{align}
b_2=\sum_i
e^{|E^i_b|/T}+\int_0^\infty\frac{dk}\pi\frac{d\delta(k)}{dk}e^{-k^2/mT}
\label{b2ori}
\end{align}
is the second virial coefficient. Here the $E^i_b$ are the bound
energies of the two-body attractive interaction $V(r)$, $\delta(k)$
is the s-wave scattering phase shift, and
 $\lambda=(2\pi/mT)^{1/2}$ is the thermal wavelength. In the BEC regime,
we consider only the relevant bound state with $E_b=-1/ma^2$.  In
the regime $T \ll 1/mr_0^2$, the values of $k$ entering
 (\ref{b2ori}) are $\ll 1/r_0$, so that
$\cot\delta(k)=-1/ka$; hence $d\delta(k)/dk=-
a/\left((ka)^2+1\right)$, and
\begin{align}
b_2
  =\left(\frac12+\frac1{\sqrt\pi}\int^{\lambda/\sqrt{2\pi}a}_0
  e^{-t^2} dt\right)e^{\lambda^2/2\pi a^2},\label{b2}
\end{align}
a dimensionless function of $\lambda/a$. Thus
\begin{equation}
C=\frac{mkT}{4\pi \Omega}\frac{\partial\log Z}{\partial a^{-1}}
=\sqrt2 n^2 \lambda^2 \frac{\partial b_2}{\partial(\lambda/a)},
\label{chigh}
\end{equation}
where we replace $z$ by the number density $n=(T/\Omega)(\partial \log Z/\partial\mu)=(z/\lambda^3)\left(1+2^{3/2}zb_2\right)$.

For $T\gg T_a\equiv 1/ma^2$ ($a\gg\lambda$), the system is a weakly
interacting Fermi gas for all coupling strengths. From  (\ref{b2}),  $\partial b_2/\partial(\lambda/a)=1/\sqrt2 \pi$, and thus
\begin{equation}
C(T)=\frac{n^2\lambda^2}{\pi} = \frac{2n^2}{mT}.
\label{HighT}
\end{equation}
Equation (\ref{HighT}) combined with (\ref{clow}) leads to the important result that the correlation $C$ has a
maximum at a \emph{nonzero} temperature $T_{max}$.

In the BCS limit, for  $T_F\ll T\ll T_a$,
$b_2= -a/\sqrt{2}\lambda$ and $C(T)=n^2a^2$; this is
simply the Hartree term in the low $T$ result (\ref{BCSlimit})
persisting to higher temperatures.
From  (\ref{BCSlimit}), (\ref{cfl}) and (\ref{HighT}), we expect that
$T_{max}\sim T_F$ and that  the relative size of the maximum, $[C(T_{max})-C(0)]/C(0)\sim |k_F a|$, is small.

In the BEC limit, when the temperature is lowered to $T\sim T_a$
($\lambda\sim a$), the bound state starts to dominate in
 (\ref{b2}) and $b_2\sim \exp(\lambda^2/2\pi a^2)$. The
correlation $C$ increases from $\sim n^2\lambda^2$ to $\sim
n^2\lambda^3\exp(\lambda^2/2\pi a^2)/a$. The virial expansion for
the Fermi gas breaks down when $b_2z\sim 1$, which happens at $T\sim
T_a/\log(1/k_Fa)$.  At this point the number of the bosonic
molecules becomes significant. Within the
regime $T_F\ll T \ll T_a/\log(1/k_Fa)$, a virial expansion
can be carried out for the bosonic molecules. The
calculation proceeds as for the fermions except that there is
no bound state for the molecules and one has to add the binding
energy $-n/ma^2$ to the free energy density, yielding
\begin{equation}
C(T)=\frac{n}{2\pi
a}\left[1+\frac{(k_Fa)^3}{6\pi}\left(\frac{\partial a_m}{\partial
a}\right)\right]. \label{BECvirial}
\end{equation}
The second term in  (\ref{BECvirial}) comes from the mean field
energy for a normal weakly interacting Bose gas, which is twice that
of a condensed Bose gas [cf.~(\ref{BEClimit})]. From  (\ref{BEClimit}), (\ref{HighT}) and (\ref{BECvirial}),
we expect that $C$ reaches a maximum at $T_{max}\sim T_c\sim T_F$, and that $[C(T_{max})-C(0)]/C(0)\ll 1$.

In the unitarity limit, (\ref{Unitaritylimit}) and (\ref{HighT}) indicate that $C$ has a maximum at
$T_{max}\sim T_F$  with  $[C(T_{max})-C(0)]/C(0)\sim1$. The pronounced maximum
in the correlations in the strong coupling limit indicates the importance
of phonons there.

\begin{figure}
 \includegraphics[width=1.0\columnwidth]{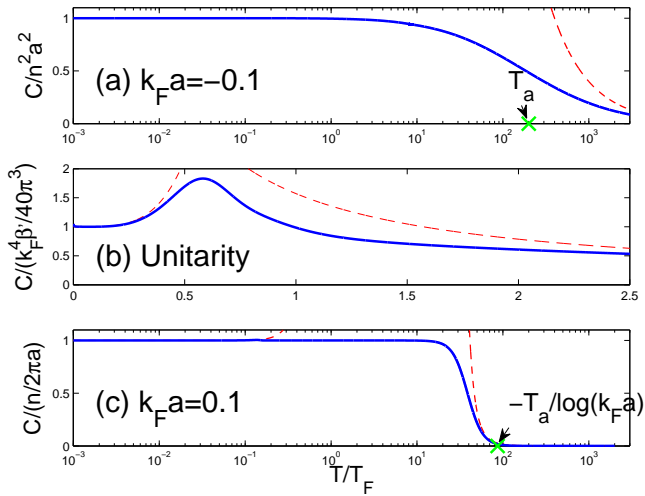}
\caption{(color online) The correlation strength $C(T)$ in the  BCS (a),  unitary (b), and  BEC (c) limits.
The dashed lines at high $T$ are the virial results, and the dashed line at low $T$ in (b) is from (\ref{Unitaritylimit}).
The maximum in $C(T)$ for $T=T_{\rm max}$ in the BCS and BEC limits is too small to be visible here.}
 \label{Cplot}
\end{figure}

Figure \ref{Cplot} sketches $C(T)$ in the BCS, unitarity, and BEC regimes.
The solid curves are interpolations between the low $T$ results
(\ref{BCSlimit})-(\ref{Unitaritylimit}) and the virial expansion
results (\ref{chigh})-(\ref{BECvirial}) for $T\gg T_F$.
Note that $C$ is continuous at $T_c$ since
the superfluid transition is second order.

\begin{figure}[isentrope]
\includegraphics[width=8cm]{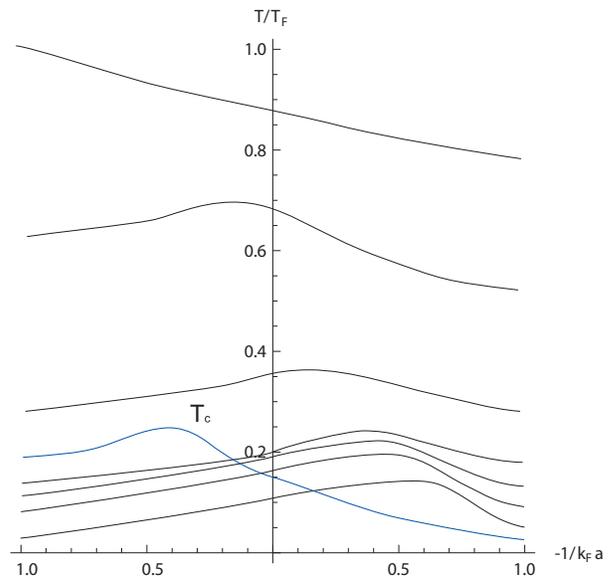}
\caption{(color online) Isentropes of a balanced two-component Fermi
gas, from which one can infer the temperature dependence of $C(T)$.}
\label{isentrope_plot}
\end{figure}

The link between the entropy and  the short range correlations
through (\ref{e:c2}) is illustrated in Fig.~(\ref{isentrope_plot}),
where we sketch the isentropes in the BEC-BCS crossover regime, as
well as the superfluid transition temperature $T_c$
\cite{dilute,boris,gmb61}. It follows from (\ref{e:c2}) and the
virial expansion result (\ref{HighT}) that the isentropes has a
negative slope in the $T,-1/k_Fa$ plane for high $T$.  Likewise, the
increase in $C(T)$ given by (\ref{BCSlimit})--(\ref{Unitaritylimit})
means that the isentropes has a postive slope for low $T$. The
correlation strength $C(T)$ has an extremum where the slope of the
isentropes as a function of $-1/k_Fa$ vanishes. Additional
information is obtained from a recent experiment, where a Fermi gas
was prepared in the BCS limit with initial energy $E_i$ and entropy
$S_i$ and then adiabatically tuned to the unitary point where the
final energy $E_f$ was measured~\cite{thomas}. The temperature was
deduced from $T=\partial E/\partial S$.  Within the experimental
regime the final temperatures $T_f$ at unitarity were generally
higher than the initial temperatures $T_i$ in the BCS limit
\cite{thomas}.  However, for the isentrope ending at
$T_f\approx0.2T_F$,  $T_i\approx T_f$, which indicates that this
isentrope (as well as neighboring ones) first bends upwards and then
downwards from the unitarity region to the BCS limit. Well below the
BCS transition, where phonons dominate the entropy, the isentropes
have positive slope, not visible on the scale of
Fig.~(\ref{isentrope_plot}).

\section{photoassociation experiment}

Photoassociation experiments provide a direct measure of the
correlation strength~\cite{randy, castin}. In a recent experiment,
$^6$Li atoms were trapped in the lowest two hyperfine states
$|1\rangle$ and $|2\rangle$. Then the bare closed channel
spin-singlet molecular state $|\psi_{v=38}\rangle$ associated with
the 834G Feshbach resonance in the 1-2 channel was excited by a
laser field $\mathbf E_L$ to a spin-singlet molecular state
$|\psi_{v'=68}\rangle$, with linewidth $\gamma=(2\pi) 11.7$ MHz.
The excited molecules were lost from the trap and the remaining
atoms counted for various durations of the laser pulse~\cite{randy}.
By Fermi's golden rule, the local loss rate of the number of atoms,
$2n$, in the trap is \cite{castin}
\begin{align}
\Gamma=-2\frac{dn(t)}{dt}=2\frac{\Omega_R^2}{\gamma}n_b,\label{rate}
\end{align}
where $\Omega_R=\langle \psi_{v'=68}|\mathbf E_L\cdot \mathbf d|\psi_{v=38}\rangle$ is the Rabi frequency,  $\mathbf d$ is the atomic dipole operator,
and $n_b$ the local density of molecules in the closed channel.
The density of closed channel molecules  $n_b$ is related to the correlation
function between the states $|1\rangle$ and $|2\rangle$ in the open channel by
\cite{castin}
\begin{align}
n_b=\frac{4\pi a_{bg} C}{m\mu_b\Delta
B}\left(\frac1{a_{bg}}-\frac1{a}\right)^2,
\label{Nb}
\end{align}
where $a_{bg}$ is the background scattering length, the molecular
magnetic moment $\mu_b$ is twice the Bohr magneton $\mu_B$, and
$\Delta B$ is the width of the Feshbach resonance.

In a homogeneous gas in the high temperature limit, we find from  (\ref{HighT}), (\ref{rate}),
and (\ref{Nb}) that $\Gamma\sim n^2/T$. In
a trap, the local density approximation gives $n(\rr)=
N\left(\bar\omega/\lambda T\right)^3\exp[-\beta V(\rr)]$ with
$V(\rr)=(m/2)(\omega_x^2 x^2+\omega_y^2 y^2+\omega_z^2 z^2)/2$ and
$\bar\omega=(\omega_x\omega_y\omega_z)^{1/3}$.  The average of $C$
over the trap, and thus the average rate
$\bar\Gamma$, is $\propto \bar\omega^3N^2/T^{5/2}$, and  Eq.~(\ref{rate}) has the solution
\begin{align}
\frac{N(t)}{N(0)}=\left(1+\frac{\Omega_R^2N_b}{\gamma N}t\right)^{-1}.
\label{HighTtrap}
\end{align}
Analogous expressions for the $T=0$ trap-averaged loss rate were
given in \cite{castin}.

\begin{figure}
\includegraphics[width=8cm]{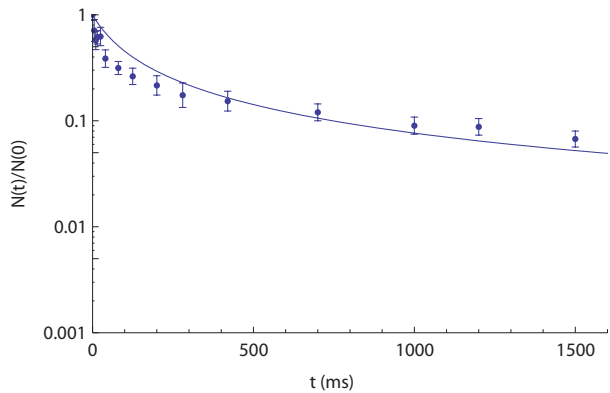}
\caption{(color online) Loss of $^6$Li atoms in the trap through
photoassociation vs. the probe duration at $B=865$ G. The solid
curve is the numerical evaluation of  (\ref{HighTtrap}). The
circles are the $T=0.75T_F$ experimental data from Ref.~\cite{randy}.}
\label{photo_bcs}
\end{figure}

In Fig.~\ref{photo_bcs}, we compare the high $T$ result
(\ref{HighTtrap}) with the experimental data obtained on the BCS
side of the resonance~\cite{randy}. Initially the temperature is
below $T_f$, where (\ref{rate}) underestimates the loss; however,
owing to depletion the Fermi temperature falls below $T$, and
(\ref{HighTtrap}) provides reasonable agreement with the data --
with no fitting parameters.

\section*{Acknowledgements}

This work grew out of the Niels Bohr International Academy Summer
Institute on Cold Atoms and Quark-Gluon Plasmas.  We are
particularly indebted to Randy Hulet for discussions there of the
Rice photoassociation experiments and providing us with the
experimental data.  This research was supported in part by NSF Grant
PHY07-01611.


\begin{thebibliography}{20}

\bibitem{mfisher}  M. Fisher, Phil. Mag. \textbf{7} 1731 (1962);  M. Fisher and J. Langer, Phys. Rev. Lett.  \textbf{20} 655 (1968).

\bibitem{bpyz} G. Baym, C. Pethick, Z. Yu and M. Zwierlein, Phys. Rev. Lett. \textbf{99}, 190407 (2007).

\bibitem{zwerger} M. Punk and W. Zwerger,  Phys. Rev. Lett. \textbf{99}, 170404 (2007).

\bibitem{zhang} S.~Zhang and A.J. Leggett, Phys. Rev. A \textbf{79}, 023601 (2009).

\bibitem{shina_ann1} S. Tan, Ann. of Phys. \textbf{323} 2952 (2008);  \textit{ibid}  \textbf{323} 2971 (2008).

\bibitem{Braaten} E. Braaten and L. Platter, Phys. Rev. Lett. \textbf{100} 205301 (2008).

\bibitem{giorgini} G.E. Astrakharchik, J. Boronat, J. Casulleras, and S. Giorgini, Phys. Rev. Lett. \textbf{93},
200404 (2004).

\bibitem{giorgini_corr} C. Lobo, I. Carusotto, S. Giorgini, A. Recati, and S. Stringari, Phys. Rev. Lett. \textbf{97}
100405 (2006).

\bibitem{randy} G. B. Partridge, K. E. Strecker, R. I. Kamar, M. W. Jack, and R. G.
Hulet, Phys. Rev. Lett. \textbf{95}, 020404 (2005).

\bibitem{castin} F. Werner, L. Tarruell, and Y. Castin,
Euro. Phys. J. B (online, 2009),
DOI: 10.1140/epjb/e2009-00040-8; arXiv:0807.0078.

\bibitem{leggettbook} A.J. Leggett, \textit{Quantum liquids: Bose condensation and Cooper pairing in condensed-matter systems}, (Oxford Univ. Press, Oxford, 2006).

\bibitem{huang} K. Huang and C.N. Yang, Phys. Rev. \textbf{105}, 767 (1957).

\bibitem{lee} T.D. Lee and C.N. Yang, Phys. Rev. \textbf{105}, 1119 (1957).

\bibitem{galitskii} V. M. Galitski\v \i , JETP, \textbf{34}:151 (1958), 1011 [Engl. transl. Soviet Phys. JETP, \textbf{7}:104 (1958), 698].

\bibitem{petrov} D.S. Petrov, C. Salomon, and G.V. Shlyapnikov, Phys. Rev.
Lett. \textbf{93}, 090404 (2004); I.V. Brodsky, M.Y. Kagan, A.V.
Klaptsov, R. Combescot, X. Leyronas, Phys. Rev. A \textbf{73},
032724 (2006).

\bibitem{sound} J. Joseph, B. Clancy, L. Luo, J. Kinast, A. Turlapov, and J.E. Thomas, Phys. Rev. Lett. {\textbf 98}, 170401 (2007).

\bibitem{Combescot} R.\ Combescot, M.\ Y.\ Kagan, and S.\ Stringari, Phys. Rev. A \textbf{74}, 042717 (2006).

\bibitem{dilute}
G. Baym, J.-P. Blaizot, M. Holzmann, F. Lalo\"e, and D. Vautherin,
Phys. Rev. Lett. \textbf{83}, 1703 (1999).

\bibitem{boris}
E. Burovski, E. Kozik, N. Prokofev, B. Svistunov, and M. Troyer,
Phys. Rev. Lett. \textbf{101}, 090402 (2008)

\bibitem{gmb61} L.P. Gor'kov and T.K. Melik-Barkhudarov, Sov. Phys. JETP {\textbf 13}, 1081 (1961).

\bibitem{thomas}L. Luo and J.E. Thomas, J. Low Temp. Phys. {\textbf 154}, 1 (2009).

\bibitem{PethickSmith} C. J. Pethick and H. Smith, \textit{Bose-Einstein condensation in dilute gases}, 2nd ed. (Cambridge Univ. Press, Cambridge, 2008).

\bibitem{landau} L.D. Landau and E.M. Lifshitz, \textit{Statistical physics I} (Pergamon Press, Oxford, 1980).

\bibitem{jason} T-L Ho and E.J. Mueller, Phys. Rev. Lett. \textbf{92}, 160404 (2004).

\bibitem{gordon_chris} G. Baym and C.J. Pethick, \textit{Landau Fermi-liquid theory} (John Wiley $\&$ Sons, Inc, New York, 1991).

\bibitem{lifshitz} E.M. Lifshitz and L.P. Pitaevskii, \textit{Statistial physics II} (Pergamon Press, Oxford, 1980).

\end{thebibliography}
\end{document}